\documentclass{ws-ijmpa}
\usepackage{bbm}
\usepackage{graphicx}
\usepackage{epsfig}
\usepackage{bm}
\usepackage{amsfonts}
\usepackage{epstopdf}
\usepackage{multirow}
\usepackage[dvipdfm,colorlinks,linkcolor=red,anchorcolor=blue,citecolor=green]{hyperref}
\usepackage[super,compress]{cite}
\usepackage{subfigure}
\renewcommand{\theequation}{\thesection.\arabic{equation}}
\newcommand\diff{\mathrm{d}}
\newcommand\Diff{\mathrm{D}}
\newcommand\im{\mathrm{i}}
\newcommand\e{\mathrm{e}}
\begin{document}

\title{Thermal Casimir
Effect for Rectangular Cavities inside D+1-dimensional Minkowski Spacetime Revisited}

\author{Rui-Hui Lin}
\author{Xiang-Hua Zhai\\
	}

\address{Shanghai United Center for Astrophysics (SUCA), Shanghai Normal University,\\
    100 Guilin Road, Shanghai 200234, China\\
    zhaixh@shnu.edu.cn
    }

\maketitle

\begin{abstract}
We reconsider the thermal scalar Casimir effect for $p$-dimensional rectangular cavity inside $D+1$-dimensional Minkowski space-time.
In dealing with the thermal Casimir effect,
one can get the free energy consisting of the divergent zero-temperature part
and the finite temperature-dependent part after some standard treatment of the finite temperature quantum field theory.
Due to the finiteness, there exists some misunderstanding about the regularization of the temperature-dependent part.
In fact, the Casimir free energy should exhibit classical limit at high temperature
where the result does not depend on the Planck constant
and the Casimir force should tend to zero with the separation increasing to infinity.
These can only be achieved after the regularization for both parts of the free energy.
We derive rigorously the regularization of the temperature-dependent part of the free energy by making use of the Abel-Plana formula repeatedly
and get the explicit expression of the terms to be subtracted.
In the cases of $D$=3, $p$=1 and $D$=3, $p$=3,
we precisely recover the results of parallel plates and three-dimensional box in the literature.
Furthermore, for $D>p$ and $D=p$ cases with periodic,
Dirichlet and Neumann boundary conditions,
we give the explicit expressions of the Casimir free energy in both low temperature (small separations) and high temperature (large separations) regimes,
through which the asymptotic behavior of the free energy changing with temperature and the side length is easy to see.
We find that for $D>p$, with the side length going to infinity,
the Casimir free energy tends to positive or negative constants or zero, depending on the boundary conditions.
But for $D=p$, the leading term of the Casimir free energy for all three boundary conditions is a logarithmic function of the side length.
We also discuss the thermal Casimir force changing with temperature and the side length in different cases
and find with the side length going to infinity the force always tends to zero for different boundary conditions regardless of $D>p$ or $D=p$.
The Casimir free energy and force at high temperature limit behave asymptotically alike in that they are proportional to the temperature,
be they positive (repulsive) or negative (attractive) in different cases.
Our study may be helpful in providing a comprehensive and complete understanding of this old problem.

\keywords{Casimir effect; finite temperature; zeta-function regularization; Abel-Plana formula.}

\end{abstract}
\ccode{PACS numbers: 03.70.+k, 11.10.-z}

\section{Introduction}
\label{intro}
The Casimir effect, as one of the most direct manifestations of the existence of vacuum oscillations, after the development for more than 60 years, is
still an active research area attracting increasing interest in both fundamental and applied
science. There are many articles and books
that can give a comprehensive review towards this
effect\cite{bordag1,bordag2,plunien,milonni,mostepanenko,krech,milton}. The influence of
temperature on the Casimir effect is a topic which should be paid
attention to since the typical quantum state is a state containing
particles in thermal equilibrium and actually all experimental
measurements of the Casimir effect have been done at room
temperature. In fact, the topic has indeed attracted a lot of
interest. In the work of Lifshitz and collaborators,
\cite{lifshitz,dzyaloshinskii} who generalized Casimir's result to
the case of parallel dielectrics, they considered temperature
corrections. And the further study were done by Sauer, \cite{sauer}
Mehra\cite{mehra} and Brown and Maclay\cite{brown}. Some years
later, the more complete theory was given by Schwinger et. al.
\cite{schwinger}. In recent years, the interest in the study on
temperature dependence of the Casimir effect is increasing. The finite temperature Casimir effect in the presence of extra dimensions\cite{brevik2} is one of the topics.  Some controversies were discussed (See Ref. \cite{milton2,brevik,milton3}
and references therein.). The progress is exciting in that there
is a possibility to measure the thermal effect in the Casimir force
\cite{geyer,klimchitskaya,klimchitskaya2}.

In the study of the Casimir effect, the key step is to
regularize the sum of infinite series to extract the physical finite
quantity. The frequently used regularization methods, such as the
Green's function method, the Abel-Plana formula and the zeta function technique,
developed for calculation of the relevant Casimir energy at zero
temperature, can be generalized straightforwardly to the finite
temperature case\cite{elizalde1,elizalde2,kirsten1,kirsten2}. With these
regularization methods, the Casimir energy at finite temperature has
been calculated for various fields and different geometries in more
than four dimensional spacetime\cite{bordag1}. The earliest work
about the Casimir effect at finite temperature
for a $p$-dimensional rectangular cavity inside a $D$-dimensional
space was done by Ambj{\o}rn and Wolfram\cite{ambjorn}. Recently
Lim and Teo \cite{lim} reconsidered this topic in detail, giving the results
of different boundary conditions and the low temperature and high temperature expansions. Later, Geyer et. al gave a critical
discussion on the thermal Casimir effect in ideal metal rectangular
boxes in three-dimensional space \cite{geyer2} where they obtained
the finite expression for the Casimir free energy by making use of
the zeta function technique. They pointed out a
problem existed in some previous papers that the subtraction of the
contribution of the black body radiation
and other geometrical contributions of quantum origin were neglected, which led to
the contradiction with the classical limit in high temperature
regime. For the finite temperature Casimir effect between the
simplest configuration -- two parallel perfectly conducting plates
in three-dimensional space, from which the rectangular boxes extend
easily, there is a standard result in literature
\cite{bordag2,mostepanenko,milton,brown} which was obtained usually
by Green's function regularization. Geyer et. al. \cite{geyer2}
reproduced the standard result by making use of the zeta function
regularization technique. More recently, Bezerra et. al.\cite{bezerra}
reconsidered the thermal Casimir effect in Einstein and closed Friedmann
universes and emphasized the same procedure as what Geyer et. al. did.
In the two new works \cite{teo1,teo2}, Teo considered the Casimir effect on the boundaries of $D$-dimensional cavities and spherical shells
and discussed the high temperature asymptotics.
In these papers \cite{geyer2,bezerra,teo1,teo2},
a common recognition was reached that the terms of order equal to or more than the square of the temperature should be subtracted but the reason was not stated clearly.
Furthermore, it is not easy to get the explicit expression of these terms from the calculation of the heat kernel coefficients.

In this paper, we reconsider systematically the thermal Casimir effect for massless scalar fields in $p$-dimensional rectangular cavity of $D+1$-dimensional Minkowski space-time.
For simplicity, we take the cavity as a hypercube.
As is known to all, the free energy can be divided into zero temperature part and temperature-dependent part.
For a restricted spatial volume,
the zero temperature part is divergent while the temperature-dependent one is finite.
So, there is no doubt on the regularization of the zero temperature part,
but due to the finiteness, there exists some misunderstanding in the literature about the regularization of the temperature-dependent part.
In fact, the Casimir free energy should exhibit classical limit at high temperature where the result does not depend on the Planck constant \cite{revzen,feinberg,rubin}
and the Casimir force should tend to zero with the separation increasing to infinity.
However, this can only be achieved after the regularization of both parts of the free energy.
In this paper, we give a rigorous prove and derivation of the regularization of the free energy of the $p$-dimensional hypercube inside $D+1$-dimensional space-time
for all three boundary conditions (BCs) of the scalar field.
The regularization of the zero temperature energy can be undoubtedly done using for instance the zeta-function technique.
For the temperature-dependent part,
we do the regularization using the Abel-Plana formula repeatedly
and obtain the explicit expression of the terms that should be subtracted from the free energy.
In the case of periodic BCs,
there is only one term to be subtracted which is corresponding to the black body radiation.
In the cases of Dirichlet and Neumann BCs,
the terms to be subtracted are expressed in series form that are proportional to the power of the side length of the hypercube and the temperature.
This result is general that the the results of the cases of parallel plates and three-dimensional box in the literature could be included in.
Furthermore, for $D>p$ and $D=p$ cases with periodic, Dirichlet and Neumann BCs,
we give the explicit expressions of the Casimir free energy in both low temperature (small separations) and high temperature (large separations) regimes,
through which asymptotic behavior of the Casimir free energy changing with temperature and the side length is easily to see.
We also discuss the thermal Casimir force changing with temperature and the side length in different cases. We use the natural units
$\hbar=c=k_\text{B}=1$ in this paper.

The structure of the paper is as follows.
In Sect. \ref{low} the free energy is divided into zero temperature part and temperature-dependent part for three kinds of BCs,
where the zero point energy is regularized and the temperature-dependent part is to be regularized.
In Sect. \ref{temperaturereg} we do the regularization of the temperature-dependent part using Abel-Plana formula
and obtain the general results indicating the subtraction of the terms proportional to the powers of the side length to get the physical result.
Sect. \ref{high} contains the results of the physical Casimir free energy in high temperature (large side length) regime.
Sect. \ref{d=p} is the consideration of $D=p$ case that is something different from the result of $D>p$ case.
The numerical computation is given in Sect. \ref{numerical} and Sect. \ref{conclusion} includes the conclusions and discussion.

\section{Two parts of the free energy}
\label{low}
In quantum field theory, the easiest and most frequently used method to treat a system at thermal
equilibrium at temperature $T$ is the imaginary-time Mastsubara formalism in which the time coordinate
makes a rotation $t\rightarrow-\mathrm{i}\tau$ and the Euclidean time $\tau$ is confined to the interval
$\tau$$\in$[0, $\beta$], where $\beta=1/T$. Periodic BC $\varphi(\tau+\beta,
 \textbf{x})=\varphi(\tau,\textbf{x})$ for bosonic field are imposed in the imaginary time coordinate.
 The partition function $\mathcal{Z}$ is given by

\begin{equation}\label{eq1}
	\mathcal{Z}=C\int\Diff\varphi\e^{-S_\text{E}[\varphi]},
\end{equation}
where $S_\text{E}[\varphi]$ is the Euclidean action. For massive bosonic field in $D$-dimensional space,

\begin{equation}\label{eq2}
	S_\text{E}[\varphi]=\frac{1}{2}\int_ 0^\beta\diff\tau\int\diff^D\mathbf{x} \varphi K_\text{E} \varphi,
\end{equation}
where $K_\text{E}=-\square_\text{E}+m^2$ and $\square_\text{E}=\frac{\partial^2 }{\partial
\tau^2}+\vartriangle$ is Euclidean wave operator. Using the standard calculation in quantum field theory
and dropping the irrelevant constant $C$, we can express the partition function as

\begin{equation}\label{eq3}
	\log (\mathcal{Z})=-\frac{1}{2}\mathrm{Tr}\log (K_\text{E}).
\end{equation}
Then the Helmholtz free energy can be expressed as

\begin{equation}
	F=-\frac 1 {\beta}\log(\mathcal{Z}).
\end{equation}

In $(D+1)$-dimensional spacetime, for a massless scalar field restricted in a $p$-dimensional hypercubic cavity with the size $L$,
and with the sizes of the left $(D-p)$-dimension $L_{p+1},L_{p+2},\cdots,L_D\gg L$,
when the scalar field satisfies periodic, Dirichlet and Neumann BCs,
the Helmholtz free energies have the following expressions, respectively
\begin{equation}
\begin{split}
	\mathcal{F}^{(\text{P})}=&\frac1{2\beta}(\prod_{j=p+1}^D\frac{L_j}{2\pi})\sum_{\substack{n_0\in\mathbb{Z}\\\vec{n}\in\mathbb{Z}^p}}\int_{-\infty}^\infty\diff^{D-p}\mathbf{r}\\
	 &\times\log[(\frac{2\pi n_0}{\beta})^2+(\frac{2\pi n_1}{L})^2+\cdots+(\frac{2\pi n_p}{L})^2+\mathbf{r}^2],\\
	\label{HelmP}
\end{split}
\end{equation}
\begin{equation}
\begin{split}
	\mathcal{F}^{(\text{D})}=&\frac1{2\beta}(\prod_{j=p+1}^D\frac{L_j}{\pi})\sum_{\substack{n_0\in\mathbb{Z}\\\vec{n}\in\mathbb{N}^p}}\int_{-\infty}^\infty\diff^{D-p}\mathbf{r}\\
	&\times\log[(\frac{2\pi n_0}{\beta})^2+(\frac{\pi n_1}{L})^2+\cdots+(\frac{\pi n_p}{L})^2+\mathbf{r}^2],\\
	\label{HelmD}
\end{split}
\end{equation}
\begin{equation}
\begin{split}
	\mathcal{F}^{(\text{N})}=&\frac1{2\beta}(\prod_{j=p+1}^D\frac{L_j}{\pi})\sum_{\substack{n_0\in\mathbb{Z}\\\vec{n}\in(\mathbb{N}\cup\{\vec{0}\})^p}}\int_{-\infty}^\infty\diff^{D-p}\mathbf{r}\\
	&\times\log[(\frac{2\pi n_0}{\beta})^2+(\frac{\pi n_1}{L})^2+\cdots+(\frac{\pi n_p}{L})^2+\mathbf{r}^2],
	\label{HelmN}
\end{split}
\end{equation}
where the superscripts (P),(D),(N) of $\mathcal{F}$ indicate the types of BCs. The $p$-dimensional vectors $\vec{n}$ in the summation signs denote the indexes $(n_1,n_2,\cdots,n_p)$, and $\vec{n}^2$ will be used to denote $\vec{n}^2=n_1^2+n_2^2+\cdots+n_p^2$ hereinafter.

In the following, we consider all three cases.
\subsection{The Periodic BCs}
To deal with the divergency, we firstly use zeta function regularization starting from some manipulation of the free energy density (energy per hyperarea)
\begin{equation}
	\begin{split}
	 f^{(\text{P})}\equiv&\frac{\mathcal{F}^{(\text{P})}}{\prod_{j=p+1}^DL_j}\\
	 =&-\frac1{2^{D-p}\pi^{\frac{D-p}{2}}\Gamma(\frac{D-p}{2})\beta}\sum_{\substack{n_0\in\mathbb{Z}\\\vec{n}\in\mathbb{Z}^p}}\lim_{s\rightarrow0}\frac{\partial}{\partial s}\int_0^\infty r^{D-p-1}[(\frac{2\pi n_0}{\beta})^2+\frac{4\pi^2\vec{n}^2}{L^2}+r^2]^{-s}\diff r,
	\label{densP}
\end{split}
\end{equation}
where the formulae
\begin{equation}
	\label{integrationformula}
	\int_{-\infty}^\infty f(\mathbf{r})\diff^D\mathbf{r}=\frac{2\pi^{\frac{D}{2}}}{\Gamma(\frac{D}{2})}\int_0^\infty r^{D-1}f(r)\diff r,
\end{equation}
and
\begin{equation}
	\log(x)=-\lim_{s\rightarrow0}\frac\partial{\partial s}x^{-s}
\end{equation}
are employed. Using
\begin{equation}
	\Gamma(s)\lambda^{-s}=\int_0^\infty t^{s-1}\e^{-\lambda t}\diff t,
	\label{ACofgamma}
\end{equation}
eq.\eqref{densP} becomes
\begin{equation}
	\begin{split}
	 f^{(\text{P})}=&-\frac1{2^{D-p}\pi^{\frac{D-p}{2}}\Gamma(\frac{D-p}{2})\beta}\\
	 &\times\sum_{\substack{n_0\in\mathbb{Z}\\\vec{n}\in\mathbb{Z}^p}}\lim_{s\rightarrow0}\frac{\partial}{\partial s}\int_0^\infty\frac{r^{D-p-1}}{\Gamma(s)}\int_0^\infty t^{s-1}\e^{-[(\frac{2\pi n_0}{\beta})^2+\frac{4\pi^2\vec{n}^2}{L^2}+r^2]t}\diff t\diff r.
	\label{densP1}
\end{split}
\end{equation}
Integrating over $r$ and using the equality
\begin{equation}
	\lim_{s\rightarrow0}\frac\partial{\partial s}\frac{u(s)}{\Gamma(s)}=u(0),
\end{equation}
where $u(s)$ is any regular function at $s=0$, one can get
\begin{equation}
	 f^{(\text{P})}=-\frac1{2^{D-p+1}\pi^{\frac{D-p}{2}}\beta}\sum_{\substack{n_0\in\mathbb{Z}\\\vec{n}\in\mathbb{Z}^p}}\int_0^\infty t^{-1-\frac{D-p}{2}}\e^{-[(\frac{2\pi n_0}{\beta})^2+\frac{4\pi^2\vec{n}^2}{L^2}]t}\diff t
	\label{densP2}
\end{equation}

With the Poisson summation formula
\begin{equation}
	\sum_{l=-\infty}^\infty\e^{-zl^2}=\sqrt{\frac\pi z}\sum_{k=-\infty}^\infty\e^{-\frac{\pi^2k^2}z} ,
	\label{poisson}
\end{equation}
eq.\eqref{densP2} becomes
\begin{equation}
	 f^{(\text{P})}=-\frac1{2^{D-p+1}\pi^{\frac{D-p}{2}}}\sum_{\substack{m_0\in\mathbb{Z}\\\vec{n}\in\mathbb{Z}^p}}\int_0^\infty t^{-1-\frac{D-p}{2}}\frac1{\sqrt{4\pi t}}\e^{-\frac{m_0^2\beta^2}{4t}}\e^{-\frac{4\vec{n}^2\pi^2}{L^2}t}\diff t.
	\label{densP3}
\end{equation}
Choosing the terms $m_0=0$, one can divide $f^{(\text{P})}$ into two parts.
One is the zero temperature part $\varepsilon_0^{(\text{P})}$
which can also be obtained by taking $\beta$ in eq.\eqref{densP2} to infinity and turning the summation over $n_0$ into a integral,
the other is the temperature-dependent part $f_T^{(\text{P})}$:
\begin{eqnarray}
	\label{e0P}
	\varepsilon_0^{(\text{P})}&=&-\frac1{2^{D-p+2}\pi^{\frac{D-p+1}2}}\sum_{\vec{n}\in\mathbb{Z}^p}\int_0^\infty t^{-\frac{3+D-p}2}\e^{-\frac{4\vec{n}^2\pi^2}{L^2}t}\diff t,\\
	\label{fTP}
	 f_T^{(\text{P})}&=&-\frac1{2^{D-p+1}\pi^{\frac{D-p+1}2}}\sum_{\substack{m_0\in\mathbb{N}\\\vec{n}\in\mathbb{Z}^p}}\int_0^\infty t^{-\frac{3+D-p}2}\e^{-\frac{m_0^2\beta^2}{4t}-\frac{4\vec{n}^2\pi^2}{L^2}t}\diff t.
\end{eqnarray}
By using eq.\eqref{ACofgamma} again, $\varepsilon_0^{(\text{P})}$ can be expressed as
\begin{equation}
	\varepsilon_0^{(\text{P})}=-\frac1{2L^{D-p+1}}\pi^{\frac{D-p+1}2}\Gamma(-\frac{D-p+1}2)Z_p(p-D-1),
	\label{e0P1}
\end{equation}
where the Epstein zeta function is defined as
\begin{equation}
	Z_p(s)\equiv\sum_{\vec{n}\in\mathbb{Z}^p\setminus\{\vec{0}\}}(\vec{n})^{-\frac s2}.
	\label{epstein}
\end{equation}

To remove the divergency in $\varepsilon_0^{(\text{P})}$, we use the reflection formula of $Z_p(s)$
\begin{equation}
	\pi^{-\frac s2}\Gamma(\frac s2)Z_p(s)=\pi^{\frac{s-p}2}\Gamma(\frac{p-s}2)Z_p(p-s)
	\label{reflectionofepstein}
\end{equation}
and get the convergent zero point energy density
\begin{equation}
	\varepsilon_0^{(\text{P}),\text{reg.}}=-\frac{\Gamma(\frac{D+1}2)Z_p(D+1)}{2\pi^{\frac{D+1}2}L^{D-p+1}}.
	\label{e0P2}
\end{equation}

For the temperature-dependent part $f_T^{(\text{P})}$ in eq.\eqref{fTP},
with the integral expression of the second kind modified Bessel function $K_\nu(z)$
\begin{equation}
	\int_0^\infty t^{-\nu-1}\e^{-at-\frac bt}\diff t=2(\frac ab)^{\frac{\nu}2}K_\nu(2\sqrt{ab}),
	\label{bessel1}
\end{equation}
it can be re-expressed as
\begin{equation}
	f_T^{(\text{P})}=-\frac2{(\beta L)^{\frac{D-p+1}2}}\sum_{\substack{m_0\in\mathbb{N}\\\vec{n}\in\mathbb{Z}^p}}(\frac{\sqrt{\vec{n}^2}}{m_0})^{\frac{D-p+1}2}K_{\frac{D-p+1}2}(\frac{2\pi m_0\sqrt{\vec{n}^2}\beta}L).
	\label{fTP1}
\end{equation}

From eqs.\eqref{e0P1} and \eqref{fTP1}, we know that the zero temperature part of the free energy density is divergent and we do the regularization using the reflection formula of Epstein zeta function and get the finite result $\varepsilon_0^{(\text{P}),\text{reg.}}$ in eq.\eqref{e0P2}.
But the temperature-dependent part \eqref{fTP1} is convergent for a given side length $L$.
This is the very reason that the regularization of this part was neglected in some previous papers.
In fact, the Casimir force should have such behavior that it tends to zero with the increase of the seperation.
But from eq.\eqref{fTP1},
it is not difficult to find that the free energy density and further the Casimir force density are divergent when $L$ goes large enough.
So the temperature-dependent part of the free energy density still needs to be regularized.
We will do the regularization using Abel-Plana formula repeatedly in the next section.
Now, we give the free energy density for the other two BCs.
\subsection{Dirichlet and Neumann BCs}
From eqs.\eqref{HelmD} and \eqref{HelmN}, repeating the procedure as in periodic BCs,
one can divide the free energy densities for these two BCs and get the regularized zero temperature part and the unregularized temperature-dependent part as follows,
\begin{eqnarray}
	\label{e0DN2}
	\varepsilon_0^{(\text{D/N}),\text{reg}.}&=&-\frac1{2^{D+2}L^{D-p+1}}\sum_{q=0}^{p-1}(\pm1)^q\frac{C_p^q\Gamma(\frac{D-q+1}2)}{\pi^{\frac{D-q+1}2}}Z_{p-q}(D-q+1),\\
	\label{fTDN1}
	f_T^{(\text{D/N})}&=&-\frac2{(2\beta L)^{\frac{D-p+1}2}}\sum_{\substack{m_0\in\mathbb{N}\\\vec{n}}}(\frac{\sqrt{\vec{n}^2}}{m_0})^{\frac{D-p+1}2}K_{\frac{D-p+1}2}(\frac{\pi m_0\sqrt{\vec{n}^2}\beta}L),
\end{eqnarray}
where ``$+$'' in $(\pm1)^q$ corresponds to Neumann BCs and ``$-$'' to Dirichlet ones,
and $\vec{n}\in\mathbb{N}^p$ in the summation sign for Dirichlet whereas $\vec{n}\in(\mathbb{N}\cup\{\vec{0}\})^p$ for Neumann.
Note that in eq.\eqref{e0DN2}, for Dirichlet BCs with even $p$,
there is a critical dimension of space $D_\text{crit}$ for each even $p$ to make the zero point energy density turn from positive to negative,
which was studied in detail in Ref. \citen{caruso}.
For example, for $p$=2, when $D<6$,
the zero point energy density is positive but when $D\geq 6$, it changes to negative.
Therefore, for $p=2$, $D_\text{crit}=6$.

\section{Rigorous derivation of the regularization of the temperature-dependent part}
\label{temperaturereg}
\setcounter{equation}{0}
The Abel-Plana formula
\begin{equation}
	\sum_{n=1}^\infty u(n)=-\frac12u(0)+\int_0^\infty u(x)\diff x+\im\int_0^\infty\frac{u(\im t)-u(-\im t)}{\e^{2\pi t}-1}\diff t
	\label{AbelPlana}
\end{equation}
is very useful in calculating the Casimir energy for various configurations.
If we denote the summation to be regularized as $A\equiv\sum_{n=1}^\infty u(n)$, then the regularized $A$ will be
\begin{equation}
\begin{split}
	A^{\text{reg}.}&=\sum_{n=1}^\infty u(n)-\int_0^\infty u(x)\diff x\\
	&=A-\int_0^\infty u(x)\diff x.
	\label{APreg}
\end{split}
\end{equation}
The last term in eq.\eqref{AbelPlana} is convergent in our cases.
We do not need to give explicit expression for this term in the current section, therefore,
in the following we will denote all terms related to this convergent part as $C$.
\subsection{Periodic BCs}
Now we use the Abel-Plana formula to regularize the temperature-dependent part of the free energy density in periodic BCs expressed in eq.\eqref{fTP1}.
Denoting
\begin{equation}
	g(z)\equiv-2\sum_{m_0\in\mathbb{N}}(\frac{\sqrt{z}}{m_0\beta L})^{\frac{D-p+1}2}K_{\frac{D-p+1}2}(\frac{2\pi m_0\sqrt{z}\beta}L),
	\label{g}
\end{equation}
then $f_T^{(\text{P})}=\sum_{\vec{n}\in\mathbb{Z}^p}g(\vec{n}^2)$. According to Abel-Plana formula \eqref{AbelPlana},
\begin{equation}
\begin{split}
	 f_T^{(\text{P})}=\sum_{\vec{n}\in\mathbb{Z}^p}g(\vec{n}^2)=&2\sum_{\substack{\vec{n}\in\mathbb{Z}^{p-1}\\k\in\mathbb{N}}}g(\vec{n}^2+k^2)+\sum_{\vec{n}\in\mathbb{Z}^{p-1}}g(\vec{n}^2)\\
	=&2\sum_{\vec{n}\in\mathbb{Z}^{p-1}}\Big[-\frac{1}{2}g(\vec{n}^2)+\int_0^\infty g(\vec{n}^2+x^2)\diff x\Big]+\sum_{\vec{n}\in\mathbb{Z}^{p-1}}g(\vec{n}^2)+C\\
	=&2\sum_{\vec{n}\in\mathbb{Z}^{p-1}}\int_0^\infty g(\vec{n}^2+x^2)\diff x+C\\
	=&4\sum_{\vec{n}\in\mathbb{Z}^{p-2}}\int_0^\infty g(\vec{n}^2+x_1^2+x_2^2)\diff^2x+C\\
	=&2^p\int_0^\infty g(x_1^2+\cdots+x_p^2)\diff^px+C.\\
	\label{fTPg}
\end{split}
\end{equation}
Substituting eq.\eqref{g} into eq.\eqref{fTPg}, then
\begin{equation}
\begin{split}
	f_T^{(\mathrm{P})}=&-2^{p+1}\sum_{m_0\in\mathbb{N}}\int_0^\infty(\frac{\sqrt{\vec{x}^2}}{m_0\beta L})^{\frac{D-p+1}{2}}K_{\frac{D-p+1}{2}}(\frac{2m_0\beta\pi\sqrt{\vec{x}^2}}{L})\diff^p\vec{x}+C\\
	=&-2\sum_{m_0\in\mathbb{N}}\int_{-\infty}^\infty(\frac{\sqrt{\vec{x}^2}}{m_0\beta L})^{\frac{D-p+1}{2}}K_{\frac{D-p+1}{2}}(\frac{2m_0\beta\pi\sqrt{\vec{x}^2}}{L})\diff^p\vec{x}+C\\
	=&-\frac{4\pi^{\frac{p}{2}}}{\Gamma(\frac{p}{2})}\sum_{m_0\in\mathbb{N}}\int_0^\infty x^{p-1}(\frac{\sqrt{x^2}}{m_0\beta L})^{\frac{D-p+1}{2}}K_{\frac{D-p+1}{2}}(\frac{2m_0\beta\pi\sqrt{x^2}}{L})\diff x+C\\
	=&-\frac{L^p\Gamma(\frac{D+1}{2})\zeta(D+1)}{\beta^{D+1}\pi^{\frac{D+1}{2}}}+C,
	\label{fTP2}
\end{split}
\end{equation}
where eq.\eqref{integrationformula} is used in the last second step.
Now, we can see clearly that to get the regularized result, the term has to be subtracted from $f_T^{(\text{P})}$ is
\begin{equation}
	-\frac{L^p\Gamma(\frac{D+1}{2})\zeta(D+1)}{\beta^{D+1}\pi^{\frac{D+1}{2}}}.
	\label{subtracttermP}
\end{equation}
Then, the regularized temperature-dependent part of the free energy density is
\begin{equation}
	 f_T^{(\text{P}),\text{reg}.}=f_T^{(\text{P})}+\frac{L^p\Gamma(\frac{D+1}{2})\zeta(D+1)}{\beta^{D+1}\pi^{\frac{D+1}{2}}}.
	\label{fTP3}
\end{equation}
\subsection{Dirichlet and Neumann BCs}
For Dirichlet and Neumann BCs, similarly, we denote
\begin{equation}
	h(z)\equiv-2\sum_{m_0\in\mathbb{N}}(\frac{\sqrt{z}}{2\beta Lm_0})^{\frac{D-p+1}{2}}K_{\frac{D-p+1}{2}}(\frac{m_0\pi\beta\sqrt{z}}{L}),
	\label{h}
\end{equation}
then
\begin{equation}
	f_T^{(\text{D})}=\sum_{\vec{n}\in\mathbb{N}^p}h(\vec{n}^2),\quad f_T^{(\text{N})}=\sum_{\vec{n}\in(\mathbb{N}\cup\{\vec{0}\})^p}h(\vec{n}^2).
	\label{fTDNh}
\end{equation}
According to Abel-Plana formula
\begin{equation}
\begin{split}
	 f_T^{(\text{D/N})}=\sum_{\substack{\vec{n}\in\mathbb{N}^p\\\vec{n}\in(\mathbb{N}\cup\{\vec{0}\})^p}}h(\vec{n}^2)=&\sum_{\substack{\vec{n}\in\mathbb{N}^{p-1}\\\vec{n}\in(\mathbb{N}\cup\{\vec{0}\})^{p-1}}}\sum_{k=1/0}^\infty h(\vec{n}^2+k^2)\\
	 =&\sum_{\substack{\vec{n}\in\mathbb{N}^{p-1}\\\vec{n}\in(\mathbb{N}\cup\{\vec{0}\})^{p-1}}}\Big[\pm\frac{1}{2}h(\vec{n}^2)+\int_0^\infty h(\vec{n}^2+x^2)\diff x\Big]+C,\\
	\label{fTDN2}
\end{split}
\end{equation}
where we show the summations over $\vec{n}$ for both BCs.

Now with the definition of the operators $\hat{A}$ and $\hat{B}$ as follows:
\begin{equation}
\begin{split}
	\hat{A}\:h(n_1^2+\cdots+n_q^2+z)&=\pm\frac{1}{2}h(n_1^2+\cdots+n_{q-1}^2+z),\\
	\hat{B}\:h(n_1^2+\cdots+n_q^2+z)&=\int_0^\infty h(n_1^2+\cdots+n_{q-1}^2+x^2+z)\diff x,\quad\forall q\le p,\:\forall z.
	\label{AB}
\end{split}
\end{equation}
Since $[\hat{A},\hat{B}]=0$, we have
\begin{equation}
\begin{split}
	 f_T^{(\text{D/N})}=&\sum_{\substack{\vec{n}\in\mathbb{N}^{p-1}\\\vec{n}\in(\mathbb{N}\cup\{\vec{0}\})^{p-1}}}(\hat{A}+\hat{B})h(\vec{n}^2)+C\\
	=&(\hat{A}+\hat{B})^ph(\vec{n}^2)+C\\
	=&\sum_{q=0}^pC_p^q\hat{A}^q\hat{B}^{p-q}h(\vec{n}^2)+C\\
	=&\sum_{q=0}^pC_p^q(\pm\frac{1}{2})^q\int_0^\infty h(x_1^2+\cdots+x_{p-q}^{2})\diff^{p-q}x+C\\
	=&\sum_{q=0}^{p-1}C_p^q(\pm\frac{1}{2})^q\int_0^\infty h(x_1^2+\cdots+x_{p-q}^{2})\diff^{p-q}x+(\pm\frac{1}{2})^ph(\vec{0}^2)+C\\
	=&-2\sum_{q=0}^{p-1}C_p^q(\pm\frac{1}{2})^q\sum_{m_0\in\mathbb{N}}\int_0^\infty(\frac{\sqrt{\vec{x}^2}}{2\beta Lm_0})^{\frac{D-p+1}{2}}K_{\frac{D-p+1}{2}}(\frac{m_0\pi\beta\sqrt{\vec{x}^2}}{L})\diff^{p-q}x\\
	&-2(\pm\frac{1}{2})^p\sum_{m_0\in\mathbb{N}}(\frac{\sqrt{\vec{n}^2}}{2\beta Lm_0})^{\frac{D-p+1}{2}}K_{\frac{D-p+1}{2}}(\frac{m_0\pi\beta\sqrt{\vec{n}^2}}{L})\Big\vert_{\vec{n}=\vec{0}}+C\\
	 =&-\sum_{q=0}^{p-1}(\pm1)^q\frac{C_p^qL^{p-q}\zeta(D-q+1)}{2^q\pi^{\frac{D-q+1}{2}}\beta^{D-q+1}}\Gamma(\frac{D-q+1}{2})\\
	 &-\frac{(\pm1)^p\Gamma(\frac{D-p+1}{2})}{2^p\beta^{D-p+1}\pi^{\frac{D-p+1}{2}}}\zeta(D-p+1)+C.\\
\end{split}
\end{equation}

We can see that the $q=p$ term is not of the integral to be substracted as eq.\eqref{APreg} shows,
but just the case $n_1=n_2=\cdots=n_p=0$.
Furthermore, it has nothing to do with $L$,
and hence won't contribute to the divergency of the Casimir force,
with which we are dealing in the first place.
Therefore, after the regularization,
the terms having to be subtracted are
\begin{equation}
	 -\sum_{q=0}^{p-1}(\pm1)^q\frac{C_p^qL^{p-q}\zeta(D-q+1)}{2^q\pi^{\frac{D-q+1}{2}}\beta^{D-q+1}}\Gamma(\frac{D-q+1}{2}).
	\label{subtracttermDN}
\end{equation}
So, the regularized temperature-dependent parts of the free energy densities for these two BCs are
\begin{equation}
	 f_T^{(\text{D/N}),\text{reg}.}=f_T^{(\mathrm{D/N})}+\sum_{q=0}^{p-1}(\pm1)^q\frac{C_p^qL^{p-q}\zeta(D-q+1)}{2^q\pi^{\frac{D-q+1}{2}}\beta^{D-q+1}}\Gamma(\frac{D-q+1}{2}).
	\label{fTDN3}
\end{equation}

From the regularization results we can see that in the case of periodic BCs,
there is only one term to be subtracted,
that is eq.\eqref{subtracttermP},
and in the cases of Dirichlet and Neumann BCs,
there are $p$ terms as shown in eq.\eqref{subtracttermDN}.
Let $D=p=3$, from eq.\eqref{subtracttermP}, we get
\begin{equation}
	-\frac{L^3\pi^2T^4}{90},
\end{equation}
and from eq.\eqref{subtracttermDN}, we get
\begin{equation}
	-\frac{L^3\pi^2T^4}{90},\quad\quad\mp\frac{3\zeta(3)L^2T^3}{4\pi},\quad\quad-\frac{L\pi T^2}8,
	\label{subtractterm3}
\end{equation}
where the sign ``$-$'' corresponds to Neumann BCs and the sign ``$+$'' to Dirichlet BCs.
It is obvious that the term proportional to $T^4$ is the blackbody radiation energy restricted in the volume $L^3$,
regardless of the BCs.
The three terms in \eqref{subtractterm3} for Dirichlet BCs are exactly the results obtained in the previous papers \cite{geyer2,dowker,vassilevich,nesterenko}.
Therefore, \eqref{subtracttermP} and \eqref{subtracttermDN} are the general results of the subtraction to get the physical Casimir free energy density for $p$-dimensional hypercube in $(D+1)$-dimensional spacetime.

We obtain the explicit expression of the terms to be subtracted through the regularization of using Abel-Plana formula
and the terms are indeed of order equal to or more than the square of the temperature as in Refs.\citen{geyer2,bezerra,teo1,teo2}.
But the advantage of our treatment is we get the explicit expression of these terms after the regularization.
We also emphasize that in Sect. \ref{low} we have known when the side length tends to infinity
the temperature-dependent part of the free energy density is divergent
and now after the regularization it is clear that the terms to be subtracted are exactly proportional to the powers of the side length.

At the end of this section, we give the physical free energy density for various BCs.
From eqs.\eqref{e0P2}, \eqref{fTP1}, and \eqref{fTP3},
the physical free energy density for periodic BCs is
\begin{equation}
\begin{split}
	f^{\text{(P)},\text{Phys.}}=&\varepsilon_0^{(\text{P}),\text{reg}.}+f_T^{(\text{P}),\text{reg}.}\\
	 =&-\frac{\Gamma(\frac{D+1}{2})Z_p(D+1)}{2\pi^{\frac{D+1}{2}}L^{D-p+1}}-2\sum_{\substack{m_0\in\mathbb{N}\\\vec{n}\in\mathbb{Z}^p}}(\frac{\sqrt{\vec{n}^2}}{m_0\beta L})^{\frac{D-p+1}{2}}K_{\frac{D-p+1}{2}}(\frac{2m_0\beta\pi\sqrt{\vec{n}^2}}{L})\\
	 &+\frac{L^p\zeta(D+1)\Gamma(\frac{D+1}{2})}{\pi^{\frac{D+1}{2}}\beta^{D+1}},
	\label{fPhysP}
\end{split}
\end{equation}
and from eq.\eqref{e0DN2}, \eqref{fTDN1}, and \eqref{fTDN3}, the results for Dirichlet/Neumann BCs are
\begin{equation}
\begin{split}
	f^{(\text{D/N}),\text{Phys.}}=&\varepsilon_0^{(\text{D/N}),\text{reg}.}+f_T^{(\text{D/N}),\text{reg}.}\\
	 =&-\frac{1}{2^{D+2}L^{D-p+1}}\sum_{q=0}^{p-1}\frac{C_p^q(\pm1)^q\Gamma(\frac{D-q+1}{2})}{\pi^{\frac{D-q+1}{2}}}Z_{p-q}(D-q+1)\\
	 &-2\sum_{m_0\in\mathbb{N}}\sum_{\substack{\vec{n}\in\mathbb{N}^p\\\vec{n}\in(\mathbb{N}\cup\{\vec{0}\})^p}}(\frac{\sqrt{\vec{n}^2}}{2\beta Lm_0})^{\frac{D-p+1}{2}}K_{\frac{D-p+1}{2}}(\frac{m_0\pi\beta\sqrt{\vec{n}^2}}{L})\\
	 &+\sum_{q=0}^{p-1}(\pm1)^q\frac{C_p^qL^{p-q}\zeta(D-q+1)}{2^q\pi^{\frac{D-q+1}{2}}\beta^{D-q+1}}\Gamma(\frac{D-q+1}{2}).
	\label{fPhysDN}
\end{split}
\end{equation}
\section{Alternative expressions of the Casimir free energy}
\label{high}
\setcounter{equation}{0}
In Sect. \ref{low}, in regularizing the free energy density,
we employ the Poisson summation formula \eqref{poisson} over the $n_0$ summation in eq.\eqref{densP2}
so that in the resulted expressions there exist terms like $K_\nu(\alpha\frac{\beta}L)$,
as indicated in eqs.\eqref{fTP1} and \eqref{fTDN1}.
But if we employ the Poisson summation formula over the $\vec{n}$ summations,
we will get terms like $K_{\nu'}(\alpha'\frac L{\beta})$.
The two results are equivalent for the same BC case.
The only difference lies in the converging rapidness.
Because $K_\nu(z)$ converges fast for large $z$,
$K_\nu(\alpha\frac{\beta}L)$ is better for large $\beta$ (low temperature or small separations)
and $K_{\nu'}(\alpha'\frac L{\beta})$ is better for small $\beta$ (high temperature or large separations).
So, the expressions given in Sect.s \ref{low} and \ref{temperaturereg} are better for low temperature regime
and can be called as the low temperature expansions of the Casimir free energy density.
Through the similar procedure as in Sect. \ref{low},
one can get the following high temperature expansions of the free energy densities:
\begin{equation}
\begin{split}
	f'^{\text{(P)}}=&-\frac{\Gamma(\frac{D}{2})}{2\beta L^{D-p}\pi^{\frac{D}{2}}}Z_p(D)-\frac{2L^p}{\beta}\sum_{\substack{n_0\in\mathbb{N}\\\vec{m}\in\mathbb{Z}^p\setminus\{\vec{0}\}}}(\frac{n_0}{\beta L\sqrt{\vec{m}^2}})^{\frac{D}{2}}K_{\frac{D}{2}}(\frac{2n_0\pi L}{\beta}\sqrt{\vec{m}^2})\\
	&-\frac{L^p\zeta(D+1)\Gamma(\frac{D+1}{2})}{\pi^{\frac{D+1}{2}}\beta^{D+1}},\\
	\label{f'TP1}
\end{split}
\end{equation}
\begin{equation}
\begin{split}
	 f'^{(\text{D/N})}=&-\frac{(\pm1)^p\Gamma(\frac{D-p+1}{2})}{2^p\beta^{D-p+1}\pi^{\frac{D-p+1}{2}}}\zeta(D-p+1)\\
	 &-\frac{1}{2^{D+1}\beta L^{D-p}}\sum_{q=0}^{p-1}\frac{(\pm1)^qC_p^q\Gamma(\frac{D-q}{2})}{\pi^{\frac{D-q}{2}}}Z_{p-q}(D-q)\\
	 &-\frac{1}{2^{D-1}\beta}\sum_{q=0}^{p-1}C_p^q(\pm1)^qL^{p-q}\sum_{\substack{n_0\in\mathbb{N}\\\vec{m}\in\mathbb{Z}^{p-q}\setminus\{\vec{0}\}}}(\frac{2n_0}{L\beta\sqrt{\vec{m}^2}})^{\frac{D-q}{2}}K_{\frac{D-q}{2}}(\frac{4\pi n_0L\sqrt{\vec{m}^2}}{\beta})\\
	 &-\sum_{q=0}^{p-1}(\pm1)^q\frac{C_p^qL^{p-q}\zeta(D-q+1)}{2^q\pi^{\frac{D-q+1}{2}}\beta^{D-q+1}}\Gamma(\frac{D-q+1}{2}).
	\label{f'TDN1}
\end{split}
\end{equation}

Since the two expansions of the free energy densities in low and high temperature regimes are equivalent,
finite physical results should be obtained from both expansions when $L\rightarrow\infty$.
As shown in Sect. \ref{temperaturereg}, in low temperature expansions,
we regularize the temperature-dependent part of the free energy density using Abel-Plana formula since this part is divergent as $L\rightarrow\infty$.
The regularization shows that the subtractions of eqs.\eqref{subtracttermP} and \eqref{subtracttermDN} from $f_T^{(\text{P})}$ and $f_T^{(\text{D/N})}$ result in finite physical outcomes eqs.\eqref{fPhysP} and \eqref{fPhysDN}.
Now, in high temperature expansions \eqref{f'TP1} and \eqref{f'TDN1},
it is easy to see that the divergent terms as $L\rightarrow\infty$ are the last terms of each equation.
So these terms have to be removed.
Furthermore, one can see that the terms to be removed are exactly the eqs.\eqref{subtracttermP} and \eqref{subtracttermDN}.
Therefore, we get the terms related to the regularization through two methods.
Then, the physical free energy densities in high temperature regime are expressed as
\begin{equation}
\begin{split}
	f'^{\text{(P)},\text{Phys}.}=&-\frac{\Gamma(\frac{D}{2})}{2\beta L^{D-p}\pi^{\frac{D}{2}}}Z_p(D)\\
	&-\frac{2L^p}{\beta^{\frac{D}{2}+1}L^{\frac{D}{2}}}\sum_{\substack{n_0\in\mathbb{N}\\\vec{m}\in\mathbb{Z}^p\setminus\{\vec{0}\}}}n_0^{\frac{D}{2}}(\vec{m}^2)^{-\frac{D}{4}}K_{\frac{D}{2}}(\frac{2n_0\pi L}{\beta}\sqrt{\vec{m}^2}),\\
	\label{f'PhysP}
\end{split}
\end{equation}
and
\begin{equation}
\begin{split}
	 f'^{(\text{D/N}),\text{Phys}.}=&-\frac{(\pm1)^p\Gamma(\frac{D-p+1}{2})}{2^p\beta^{D-p+1}\pi^{\frac{D-p+1}{2}}}\zeta(D-p+1)\\
	 &-\frac{1}{2^{D+1}\beta L^{D-p}}\sum_{q=0}^{p-1}\frac{(\pm1)^qC_p^q\Gamma(\frac{D-q}{2})}{\pi^{\frac{D-q}{2}}}Z_{p-q}(D-q)\\
	 &-\Big[\frac{1}{2^{D-1}\beta}\sum_{q=0}^{p-1}C_p^q(\pm1)^qL^{p-q}\\
	 &\times\sum_{\substack{n_0\in\mathbb{N}\\\vec{m}\in\mathbb{Z}^{p-q}\setminus\{\vec{0}\}}}(\frac{2n_0}{L\beta\sqrt{\vec{m}^2}})^{\frac{D-q}{2}}K_{\frac{D-q}{2}}(\frac{4\pi n_0L\sqrt{\vec{m}^2}}{\beta})\Big].
	\label{f'PhysDN}
\end{split}
\end{equation}

\section{The closed case of $D=p$}
\label{d=p}
\setcounter{equation}{0}
For the closed cavities of $D=p$, Ambj$\varnothing$rn and Wolfram \cite{ambjorn} and Lim and Teo \cite{lim} have given the results of the free energy for periodic BCs.
Here, we give explicitly both the low and high temperature expansions for all three BCs by making some modifications of the results of $D>p$ case.
\subsection{Low temperature expansion}
The physical Casimir free energy densities of $D>p$ cases in low temperature regime are shown in eqs.\eqref{fPhysP} and \eqref{fPhysDN}.
In both equations, when $\vec{n}\in\{\vec{0}\}$, the result of the second terms is (for periodic and Neumann BCs but not for Dirichlet)
\begin{equation}
	-\frac{\Gamma(\frac{D-p+1}2)\zeta(D-p+1)}{\pi^{\frac{D-p+1}2}\beta^{D-p+1}},
	\label{subtracttermd=p}
\end{equation}
which is divergent for $D=p$.
However, Ambj$\varnothing$rn and Wolfram \cite{ambjorn} have argued physically that this term is the free Bose gas result
and should not appear in the result of physical free energy of $D=p$ case.
Therefore, excluding $\vec{n}\in\{\vec{0}\}$,
the physical free energies for a closed $D=p$ cavity in low temperature expansion are
\begin{equation}
\begin{split}
	 f^{\text{(P)},\text{Phys.}}_{D=p}=&-\frac{\Gamma(\frac{D+1}{2})Z_D(D+1)}{2\pi^{\frac{D+1}{2}}L}-2\sum_{\substack{m_0\in\mathbb{N}\\\vec{n}\in\mathbb{Z}^D\setminus\{\vec{0}\}}}(\frac{\sqrt{\vec{n}^2}}{m_0\beta L})^{\frac{1}{2}}K_{\frac{1}{2}}(\frac{2m_0\beta\pi\sqrt{\vec{n}^2}}{L})\\
	 &+\frac{L^p\zeta(D+1)\Gamma(\frac{D+1}{2})}{\pi^{\frac{D+1}{2}}\beta^{D+1}},
	\label{fPhysPd=p}
\end{split}
\end{equation}
and
\begin{equation}
\begin{split}
	 f^{(\text{D/N}),\text{Phys.}}_{D=p}=&-\frac{1}{2^{D+2}L}\sum_{q=0}^{D-1}\frac{C_D^q(\pm1)^q\Gamma(\frac{D-q+1}{2})}{\pi^{\frac{D-q+1}{2}}}Z_{D-q}(D-q+1)\\
	 &-2\sum_{m_0\in\mathbb{N}}\sum_{\substack{\vec{n}\in\mathbb{N}^D\\\vec{n}\in(\mathbb{N}\cup\{\vec{0}\})^D\setminus\{\vec{0}\}}}(\frac{\sqrt{\vec{n}^2}}{2\beta Lm_0})^{\frac{1}{2}}K_{\frac{1}{2}}(\frac{m_0\pi\beta\sqrt{\vec{n}^2}}{L})\\
	 &+\sum_{q=0}^{D-1}(\pm1)^q\frac{C_D^qL^{p-q}\zeta(D-q+1)}{2^q\pi^{\frac{D-q+1}{2}}\beta^{D-q+1}}\Gamma(\frac{D-q+1}{2}).
	\label{fPhysDNd=p}
\end{split}
\end{equation}
\subsection{High temperature expansion}
Remember that we still have to exclude the mode $\vec{n}\in\{\vec{0}\}$ for periodic and Neumann BCs,
that is, to substract eq.\eqref{subtracttermd=p} from eqs.\eqref{f'PhysP} and \eqref{f'PhysDN} to obtain the physical free energies for $D=p$ case.
However, it is not that direct to get to the results as in low temperature regime.
From eq.\eqref{f'PhysP} we can see that the first term is also divergent for $D=p$.
So, we have to deal with two divergent terms now.
Epstein zeta function $Z_p(s)$ can be expressed as a summation of Riemann zeta functions plus a series of $K_\nu(z)$ that converges rapidly \cite{lim}.
\begin{equation}
	Z_p(s)=\frac2{\Gamma(\frac s2)}\sum_{j=0}^{p-1}\pi^{\frac j2}\Gamma(\frac{s-j}2)\zeta(s-j)+\frac{4\pi^{\frac s2}}{\Gamma(\frac s2)}\sum_{j=1}^{p-1}\sum_{\substack{m\in\mathbb{N}\\\vec{k}\in\mathbb{Z}^j\setminus\{\vec{0}\}}}(\frac{\sqrt{\vec{k}^2}}m)^{\frac{s-j}2}K_{\frac{s-j}2}(2\pi m\sqrt{\vec{k}^2}).
	\label{epsteinexpansion}
\end{equation}
When $s=p$, the divergency of the LHS lies only in the $j=p-1$ term of the first part of the RHS.
So, the first term of eq.\eqref{f'PhysP} can be expressed using eq.\eqref{epsteinexpansion} as
\begin{equation}
\begin{split}
	-\frac{\Gamma(\frac D2)Z_p(D)}{2\beta L^{D-p}\pi^{\frac D2}}=&-\frac{\Gamma(\frac{D-p+1}2)\zeta(D-p+1)}{\beta L^{D-p}\pi^{\frac{D-p+1}2}}-\frac1{\beta L^{D-p}}\sum_{j=0}^{p-2}\frac{\Gamma(\frac{D-j}2)\zeta(D-j)}{\pi^{\frac{D-j}2}}\\
	&-\frac2{\beta L^{D-p}}\sum_{j=1}^{p-1}\sum_{\substack{m\in\mathbb{N}\\\vec{k}\in\mathbb{Z}^j\setminus\{\vec{0}\}}}(\frac{\sqrt{\vec{k}^2}}m)^{\frac{D-j}2}K_{\frac{D-j}2}(2\pi m\sqrt{\vec{k}^2}),
	\label{divergencyP}
\end{split}
\end{equation}
where the first term of the RHS is divergent and the rest are convergent when $D\rightarrow p$.
Now , together with eq.\eqref{subtracttermd=p}, the divergency can be expressed as
\begin{equation}
	-\lim_{D\rightarrow p}\frac{\Gamma(\frac{D-p+1}2)\zeta(D-p+1)}{\pi^{\frac{D-p+1}2}\beta}[\frac1{L^{D-p}}-\frac1{\beta^{D-p}}]=-\frac1\beta\log(\frac{\beta}L),
	\label{logP}
\end{equation}
where the expansions $(\frac1{a^x}-\frac1{b^x})\vert_{x\rightarrow0}\sim x\log(\frac ba)+\mathcal{O}(x)$
and $(\zeta(x+1)x)\vert_{x\rightarrow0}\sim1+\mathcal{O}(x)$ are used.
Combining eqs.\eqref{f'PhysP}, \eqref{divergencyP} and \eqref{logP},
we obtain the free energy for periodic BCs of $D=p$ case in high temperature regime as
\begin{equation}
\begin{split}
	 f^{\text{(P)},\text{Phys}.}_{D=p}=&-\sum_{j=0}^{D-2}\frac{\Gamma(\frac{D-j}{2})\zeta(D-j)}{\beta\pi^{\frac{D-j}{2}}}-\frac{2L^{\frac{D}{2}}}{\beta^{\frac{D}{2}+1}}\sum_{\substack{n_0\in\mathbb{N}\\\vec{m}\in\mathbb{Z}^D\setminus\{\vec{0}\}}}(\frac{n_0}{\sqrt{\vec{m}^2}})^{\frac D2}K_{\frac{D}{2}}(\frac{2n_0\pi L}{\beta}\sqrt{\vec{m}^2})\\
	 &-\frac{2}{\beta}\sum_{j=1}^{D-1}\sum_{\substack{m\in\mathbb{N}\\\vec{k}\in\mathbb{Z}^j\setminus\{\vec{0}\}}}(\frac{\sqrt{\vec{k}^2}}{m})^{\frac{D-j}{2}}K_{\frac{D-j}{2}}(2\pi m\sqrt{\vec{k}^2})-\frac{1}{\beta}\ln\frac{\beta}{L}.
	\label{f'PhysPd=p}
\end{split}
\end{equation}

For Dirichlet and Neumann BCs, as $D\rightarrow p$,
the first two terms in eqs.\eqref{f'PhysDN} are both divergent.
Repeating the procedure similar to eqs.\eqref{divergencyP} to \eqref{f'PhysPd=p},
we get the physical free energy for $D=p$ case in Dirichlet and Neumann BCs as
\begin{equation}
\begin{split}
	 f^{(\text{D/N}),\text{Phys}.}_{D=p}=&-\frac{1}{2^{D-1}\beta}\sum_{q=0}^{D-1}C_D^q(\pm1)^q\sum_{\substack{n_0\in\mathbb{N}\\\vec{m}\in\mathbb{Z}^{D-q}\setminus\{\vec{0}\}}}(\frac{2n_0L}{\beta\sqrt{\vec{m}^2}})^{\frac{D-q}{2}}K_{\frac{D-q}{2}}(\frac{4\pi n_0L\sqrt{\vec{m}^2}}{\beta})\\
	 &-\frac1{2^D\beta}\sum_{q=0}^{D-2}\sum_{j=0}^{D-q-2}\frac{(\pm1)^qC_D^q}{\pi^{\frac{D-q-j}{2}}}\Gamma(\frac{D-q-j}{2})\zeta(D-q-j)\\
	 &-\frac1{2^{D-1}\beta}\sum_{q=0}^{D-2}\sum_{j=1}^{D-q-1}\sum_{\substack{m\in\mathbb{N}\\\vec{k}\in\mathbb{Z}^j\setminus\{\vec{0}\}}}(\pm1)^qC_D^q(\frac{\sqrt{\vec{k}^2}}{m})^{\frac{D-q-j}{2}}K_{\frac{D-q-j}{2}}(2m\pi\sqrt{\vec{k}^2})\\
	 &+A,
	\label{f'PhysDNd=p}
\end{split}
\end{equation}
and
\begin{equation}
	A=
	\begin{cases}
		-\frac{(-1)^D}{2^D\beta}\log(\frac{2L}{\beta})\quad\quad&\text{for Dirichlet BCs};\\
		-\frac{1-2^D}{2^D\beta}\log(\frac{2L}{\beta})\quad\quad&\text{for Neumann BCs}.
	\end{cases}
	\label{logDN}
\end{equation}

Till now, we have obtained the physical free energy (density) in both low and high temperature regimes for all three BCs of $D>p$ and $D=p$ cases
expressed in eqs.\eqref{fPhysP}, \eqref{fPhysDN}, \eqref{f'PhysP}, \eqref{f'PhysDN}, \eqref{fPhysPd=p}, \eqref{fPhysDNd=p}, and \eqref{f'PhysPd=p} - \eqref{logDN},
where it is easy to find that the absolute value of the physical free energy (density) in every case is an increasing function of temperature.

\section{Numerical computation of the Casimir free energy and force}
\label{numerical}
\setcounter{equation}{0}
\subsection{The Casimir free energy}
From the results of the physical free energy (density) in all cases,
it is easy to observe their asymptotic behaviors in high temperature limits as listed in Table \ref{asymptoticf}. In the table, we only retain the term of the highest power of $\frac1{\beta}$.
The high temperature limits of the free energy (density) for Dirichlet BCs with even $p$ and for Neumann and periodic BCs both with $D>p$ are negative
whereas they are positive for Dirichlet BCs with odd $p$ and for Neumann and periodic BCs both with $D=p$.
We also restore standard units in the table,
and find that with the definition of effective temperature,
$k_\text{B}T_\text{eff}\equiv\hbar c/L$ for periodic BCs, and $2k_\text{B}T_\text{eff}\equiv\hbar c/L$ for the other two BCs \cite{mostepanenko},
the high temperature limits of the free energy (density) in all the cases are proportional to $k_\text{B}T$ and do not depend on the Planck constant,
i.e., the classical limits are achieved.
\begin{table}[!htbp]
\renewcommand{\arraystretch}{2.5}
\tbl{High temperature limits of the free energy (density)}
{\begin{tabular}{|c|c|c|c|}
	\hline
	\hline
	\multicolumn{2}{|c|}{ }			&Natural Units	&Standard Units\\
	\hline
	\multirow{2}{*}{periodic}	&$D>p$	&$-\frac{\Gamma(\frac{D}{2})Z_p(D)}{2\pi^{\frac{D}{2}}\beta L^{D-p}}$&$-k_\text{B}T\frac{\Gamma(\frac{D}{2})Z_p(D)}{2\pi^{\frac{D}{2}} L^{D-p}}$\\[2mm]
	\cline{2-4}
	&$D=p$	&$-\frac{1}{\beta}\log(\frac{\beta}{L})$&$k_\text{B}T\log(\frac T{T_\text{eff}})$\\[2mm]
	\hline
	\multirow{2}{*}{Neumann}	&$D>p$	 &$-\frac{\Gamma(\frac{D-p+1}{2})\zeta(D-p+1)}{2^p\beta^{D-p+1}\pi^{\frac{D-p+1}{2}}}$&$-k_\text{B}T\frac{\Gamma(\frac{D-p+1}{2})\zeta(D-p+1)}{2^DL^{D-p}\pi^{\frac{D-p+1}{2}}}(\frac T{T_\text{eff}})^{D-p}$\\[2mm]
	\cline{2-4}
					&$D=p$	 &$-\frac{1-2^D}{2^D\beta}\log(\frac{2L}{\beta})$&$k_\text{B}T\frac{2^D-1}{2^D}\log(\frac{T}{T_\text{eff}})$\\[2mm]
	\hline
	\multirow{2}{*}{Dirichlet}	&$D>p$	 &$-\frac{(-1)^p\Gamma(\frac{D-p+1}{2})\zeta(D-p+1)}{2^p\beta^{D-p+1}\pi^{\frac{D-p+1}{2}}}$&$-k_\text{B}T\frac{(-1)^p\Gamma(\frac{D-p+1}{2})\zeta(D-p+1)}{2^DL^{D-p}\pi^{\frac{D-p+1}{2}}}(\frac T{T_\text{eff}})^{D-p}$\\[2mm]
	\cline{2-4}
					&$D=p$	&$-\frac{(-1)^D}{2^D\beta}\log(\frac{2L}{\beta})$	 &$-k_\text{B}T(-\frac12)^D\log(\frac{T}{T_\text{eff}})$\\[2mm]
	\hline
\end{tabular}
\label{asymptoticf}}
\end{table}

\begin{figure}[htp!]
\centering
\subfigure[]{
\label{p2d5DenergyT}
\includegraphics[width=0.3\textwidth]{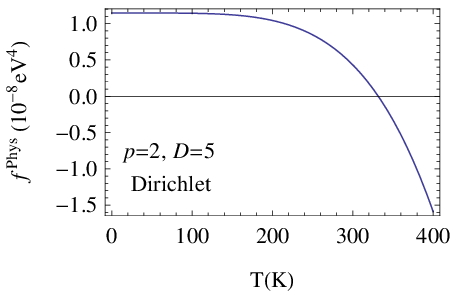}}
\subfigure[]{
\label{p2d6DenergyT}
\includegraphics[width=0.3\textwidth]{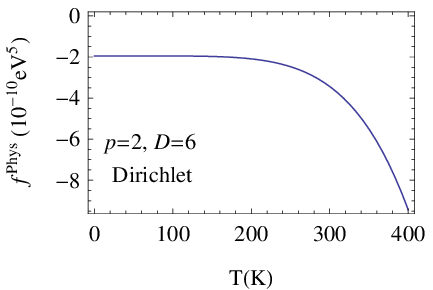}}
\subfigure[]{
\label{p3d5DenergyT}
\includegraphics[width=0.3\textwidth]{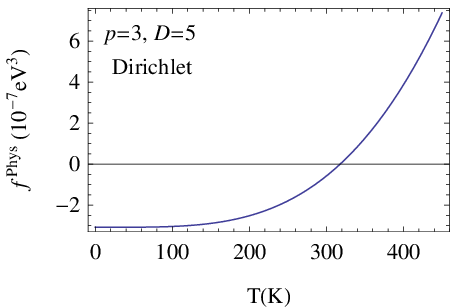}}
\caption{The free energy density as a function of temperature at the side length $L=10\mathrm{eV}^{-1}$:
\textbf{(a)} for Dirichlet BCs with $p=2$, $D=5$;
\textbf{(b)} for Dirichlet BCs with $p=2$, $D=6$;
\textbf{(c)} for Dirichlet BCs with $p=3$, $D=5$.}
\label{fofT}
\end{figure}

In Fig. \ref{fofT},
we plot the free energy density as a function of temperature at $L=10\mathrm{eV}^{-1}$ in various cases.
Fig. \ref{p2d5DenergyT} and \ref{p2d6DenergyT} are plotted for Dirichlet BCs with $p=2$, $D=5$ and $p=2$, $D=6$, respectively.
They represent the situations of even $p$, $D<D_{\text{crit}}$, and even $p$, $D\ge D_{\text{crit}}$,
recalling that for $p=2$, $D=6$ is the critical dimension for the zero point energy to change from positive to negative \cite{caruso}.
The free energy densities of Neumann and periodic BCs both with $D>p$ have the similar behavior as in Fig. \ref{p2d6DenergyT} when plotted as functions of temperature.
Fig. \ref{p3d5DenergyT} is the behavior of the free energy density for Dirichlet BCs with $p=3$, $D=5$.
In fact, the free energies (densities) for Dirichlet BCs with any odd $p$ and for Neumann and periodic BCs both with $D=p$ have similar behaviors as shown in Fig. \ref{p3d5DenergyT}.
With the increase of temperature, they change from negative to positive and increase with $T$.

\begin{figure}[htp!]
\centering
\subfigure[]{
\label{p2d5DenergyL}
\includegraphics[width=0.3\textwidth]{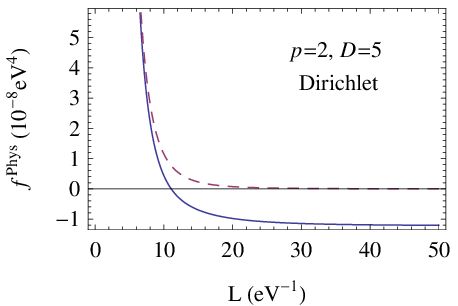}}
\subfigure[]{
\label{p2d6DenergyL}
\includegraphics[width=0.3\textwidth]{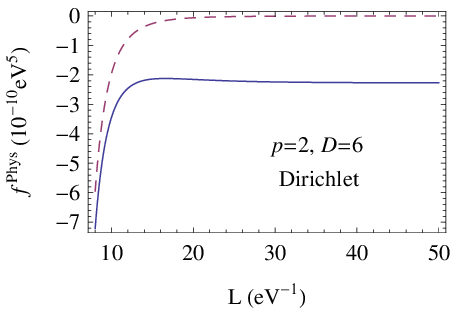}}
\subfigure[]{
\label{p2d7DenergyL}
\includegraphics[width=0.3\textwidth]{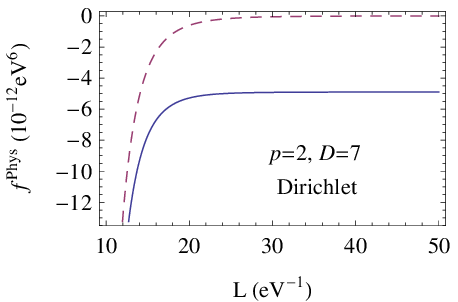}}
\subfigure[]{
\label{p3d5DenergyL}
\includegraphics[width=0.3\textwidth]{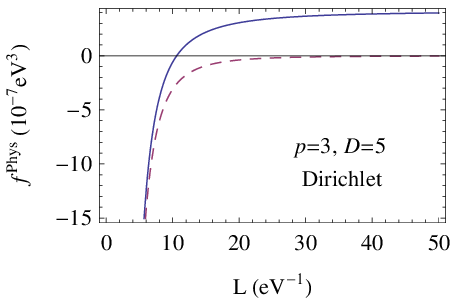}}
\subfigure[]{
\label{p3d5PenergyL}
\includegraphics[width=0.3\textwidth]{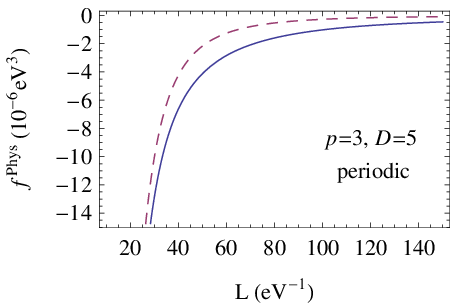}}
\caption{The free energy density as a function of the side length for $D>p$ cases.
The solid lines are the free energy densities of $T=300K$ and the dashed lines are the results of $T=0K$.
\textbf{(a)} is the case of even $p$ and $D<D_{\text{crit}}$ for Dirichlet BCs specifying $p=2$, $D=5$;
\textbf{(b)} even $p$ and $D=D_{\text{crit}}$ specifying $p=2$, $D=6$;
\textbf{(c)} even $p$ and $D>D_{\text{crit}}$ specifying $p=2$, $D=7$;
and \textbf{(d)} odd $p$ specifying $p=3$, $D=5$.
\textbf{(e)} is the case of $D>p$ for periodic BCs specifying $p=3$, $D=5$.
The behavior of the free energy density in Neumann BCs for $D>p$ is similar to what is shown in \textbf{(c)}.}
\label{fofL}
\end{figure}

Then, we study the behavior of the free energy density as a function of the side length $L$.
Note that the expressions of the free energy density for low and high temperature regimes are also good to describe small and large side lengths regimes, respectively.
It is not difficult to see the asymptotic behaviors of the free energy densities with the change of the side length from the expressions.
For example, from eqs.\eqref{f'PhysP} and \eqref{f'PhysDN},
it is obvious that for a given $T$, when $L\rightarrow\infty$, the free energy density tends to zero for periodic BCs whereas for Dirichlet and Neumann BCs it tends to a constant depending on the temperature.
These analyses are plotted in Fig. \ref{fofL}.

From Figs. \ref{fofT} and \ref{fofL},
it is obvious that for Dirichlet BCs
with even $p$ and $D<D_{\text{crit}}$,
the free energy (density) crosses the zero value and turn from positive to negative with the increase of temperature or the side length,
whereas for Dirichlet BCs with odd $p$ and for Neumann and periodic BCs both with $D=p$ it turns from negative to positive.
For other cases, that is, for Dirichlet BCs with even $p$ and $D\geq D_\text{crit}$ and for Neumann and periodic BCs both with $D>p$
there is no cross of the free energy density.
In addition, it is worth noting that there is a local maximum in Fig. \ref{p2d6DenergyL}.
We will explain it later when discussing the behavior of the Casimir force density.

\begin{figure}[htp!]
\centering
\subfigure[]{
\label{p2d2DenergyL}
\includegraphics[width=0.35\textwidth]{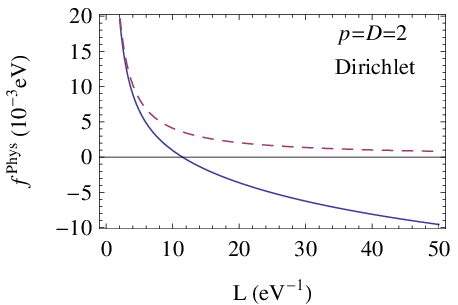}}
\subfigure[]{
\label{p3d3DenergyL}
\includegraphics[width=0.35\textwidth]{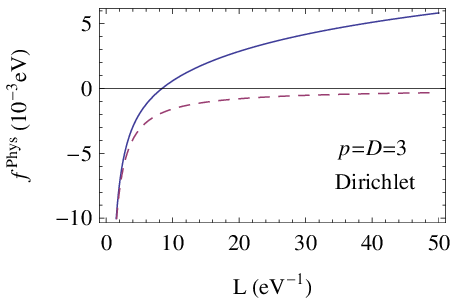}}
\caption{The free energy as a function of the side length for $D=p$ cases.
The solid lines are the free energies of $T=300K$ and the dashed lines are the reults of $T=0K$.
\textbf{(a)} is the cases of $D=p=$even of Dirichlet BCs specifying $D=p=2$.
\textbf{(b)} is the cases $D=p=$odd of Dirichlet or any $D=p$ cases of Neumann and periodic BCs, specifying $D=p=3$ in Dirichlet.}
\label{fofLd=p}
\end{figure}

For the case of $D=p$,
the situation is something different.
From eqs.\eqref{f'PhysPd=p} and \eqref{f'PhysDNd=p} together with \eqref{logDN},
for a given $T$, when $L\rightarrow\infty$,
the leading term is the last term of each equation,
which is a logarithm function of $L$,
but not a constant as in the cases of $D>p$.
In Ref. \citen{geyer2}, Geyer et. al. plotted the free energy in a cube of $D=p=3$ as a function of the side length at $T=300K$,
and claimed it approaches to a constant but not shown in their figure.
From the analytical analysis, although the low temperature and high temperature expressions of the free energy are equivalent,
the high temperature expression is necessary in that the asymptotic behavior at high temperature (large separation) can be seen directly.
The numerical results in Fig. \ref{fofLd=p} are consistent with the analytical results in eqs.\eqref{f'PhysPd=p} and \eqref{f'PhysDNd=p} and \eqref{logDN}
where the logarithmic asymptote is obvious.
However, in Ref. \citen{geyer2}, the authors did not give the high temperature expression.
Furthermore, we can not tell the difference between their Fig. 1(a) and our Fig. \ref{p3d3DenergyL}.
Therefore, we think in the case of $D=p$, at large side length,
the Casimir free energy changes as a logarithmic function of $L$ but does not approach to a constant.
It is obvious that the asymptotic behaviors shown in Figs. \ref{fofT}-\ref{fofLd=p} are in accordance with what Table \ref{asymptoticf} tells.

\begin{figure}[htp!]
\centering
\subfigure[]{
\label{d246DenergyL}
\includegraphics[width=0.3\textwidth]{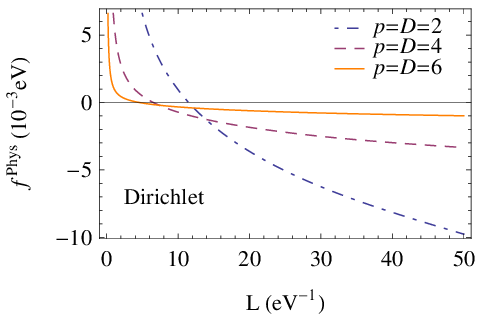}}
\subfigure[]{
\label{d357DenergyL}
\includegraphics[width=0.3\textwidth]{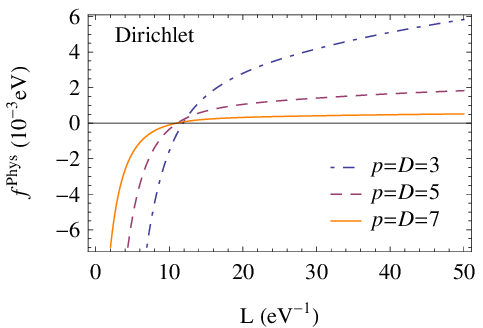}}
\subfigure[]{
\label{d345NenergyL}
\includegraphics[width=0.3\textwidth]{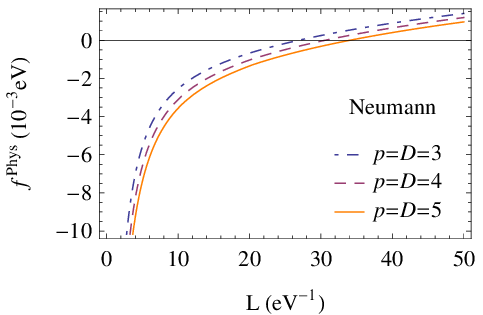}}
\caption{The free energy as a function of the side length for different $D$.
\textbf{(a)} plots the cases $p=D=2,4,6$ for Dirichlet BCs with dashed-dot, dashed and solid lines respectively, representing any cases of $D=p=$even for Dirichlet BCs;
and \textbf{(b)} the cases $p=D=3,5,7$ for Dirichlet BCs, representing any cases of $D=p=$odd.
\textbf{(c)} shows the cases $p=D=3,4,5$ for Neumann BCs, resembling the cases of periodic BCs.}
\label{fofd}
\end{figure}

We also consider the influence of the dimension on the free energy.
For $D=p$ cases of each BCs,
we plot in Fig. \ref{fofd} the free energy as a function of $L$ for different $D$.
We find that the influences of $D$ on the free energy vary slightly for different cases
but at large side length regime, the free energy decreases with the increase of $D$ for all cases.

\subsection{The Casimir force}
From the relation of the Casimir force (density) $F^{\text{Cas}.}$ to the physical free energy (density) $f^{\text{Phys}.}$
\[F^{\text{Cas}.}=-\frac{\partial f^{\text{Phys}.}}{\partial L},\]
one can easily get the Casimir force (density) in various cases considered above.
We ignore writing the expressions of the Casimir force (density) but give the numerical results as follows.

\begin{figure}[htp!]
\centering
\subfigure[]{
\label{p2d3NforceT}
\includegraphics[width=0.35\textwidth]{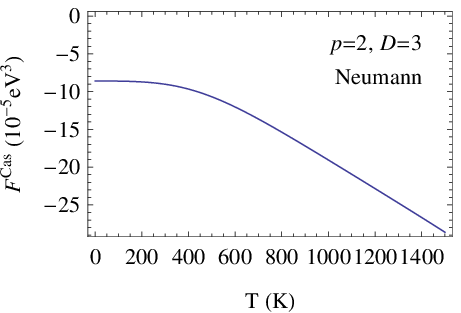}}
\subfigure[]{
\label{p2d3NforceL}
\includegraphics[width=0.34\textwidth]{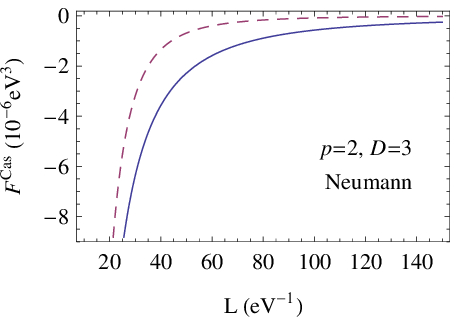}}
\caption{The Casimir force as functions of temperature and the side length with $p=2$, $D=3$ for Neumann BCs.
	In \textbf{(a)}, we take the side length as $L=10\mathrm{eV}^{-1}$,
	and in \textbf{(b)}, the dashed line is zero temperature Casimir force density
and the solid line is thermal Casimir force density at $T=300K$.}
\label{FofN}
\end{figure}

Fig. \ref{FofN} shows the Casimir force density as functions of temperature and the side length for Neumann BCs. In most of the cases including Dirichlet BCs with odd $p$, Neumann and periodic BCs, the Casimir force is attractive and it has the similar behaviors as what is shown in Fig. \ref{FofN}.

\begin{figure}[htp!]
\centering
\subfigure[]{
\label{p2d5DforceT}
\includegraphics[width=0.35\textwidth]{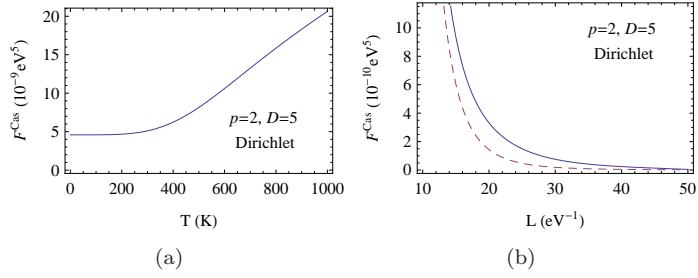}}
\subfigure[]{
\label{p2d5DforceL}
\includegraphics[width=0.35\textwidth]{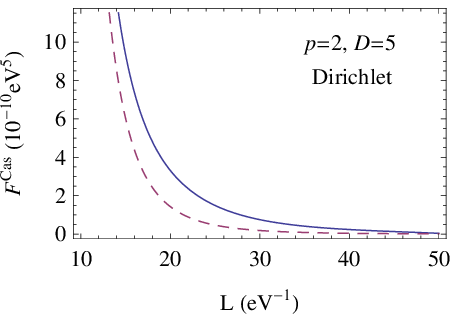}}
\caption{The Casimir force as functions of temperature and the side length for Dirichlet BCs with even $p$ and $D<D_{\text{crit}}$.
	In \textbf{(a)}, we take the side length as $L=10\mathrm{eV}^{-1}$,
	and in \textbf{(b)}, the dashed line is zero temperature Casimir force density
and the solid line is thermal Casimir force density with $T=300K$.}
\label{FofD1}
\end{figure}

\begin{figure}[htp!]
\centering
\subfigure[]{
\label{p2d6DforceT}
\includegraphics[width=0.35\textwidth]{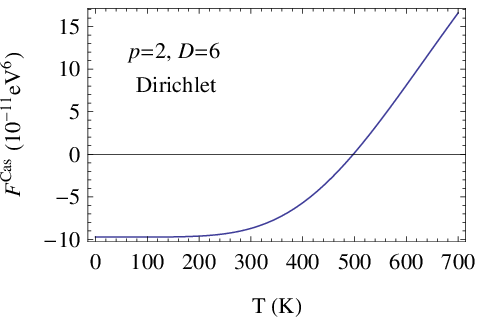}}
\subfigure[]{
\label{p2d6DforceL}
\includegraphics[width=0.35\textwidth]{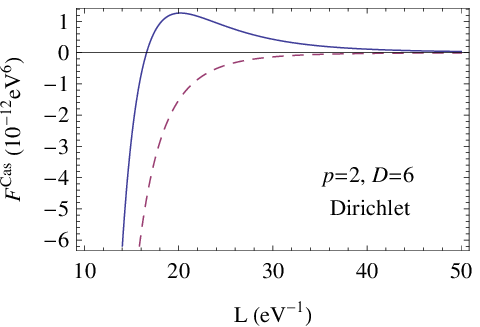}}
\caption{The Casimir force as functions of temperature and the side length for Dirichlet BCs with even $p$ and $D=D_{\text{crit}}$.
	In \textbf{(a)}, we take the side length as $L=10\mathrm{eV}^{-1}$,
	and in \textbf{(b)}, the dashed line is zero temperature Casimir force density
and the solid line is thermal Casimir force density with $T=300K$.}
\label{FofD2}
\end{figure}

\begin{figure}[htp!]
\centering
\subfigure[]{
\label{p2d7DforceT}
\includegraphics[width=0.3\textwidth]{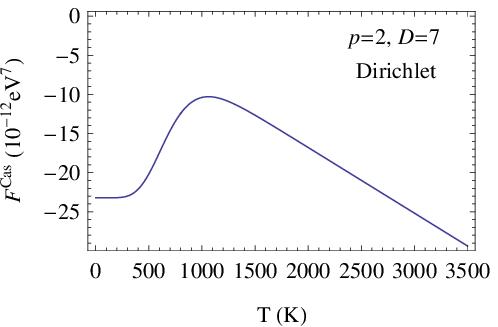}}
\subfigure[]{
\label{p2d7DforceL1}
\includegraphics[width=0.3\textwidth]{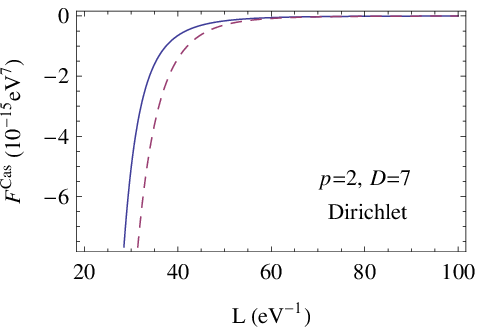}}
\subfigure[]{
\label{p2d7DforceL2}
\includegraphics[width=0.3\textwidth]{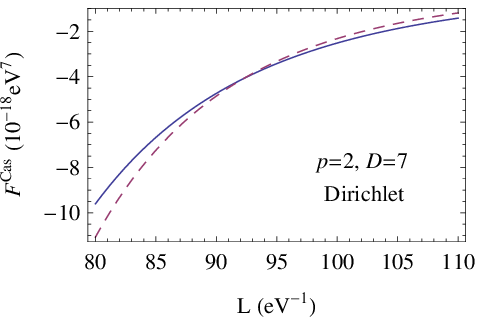}}
\caption{The Casimir force as functions of temperature and the side length for Dirichlet BCs with even $p$ and $D>D_{\text{crit}}$.
	In \textbf{(a)}, we take the side length as $L=10\mathrm{eV}^{-1}$,
	in \textbf{(b)} and \textbf{(c)}, the dashed lines are zero temperature Casimir force densities
and the solid lines are thermal Casimir force densities with $T=300K$.
\textbf{(c)} shows explicitly the cross of the thermal line and the zero temperature line.}
\label{FofD3}
\end{figure}

For the case of Dirichlet BCs with even $p$,
the situations are more complicated.
Fig. \ref{FofD1} - \ref{FofD3} show the cases of $D<D_{\text{crit}}$, $D=D_{\text{crit}}$ and $D>D_{\text{crit}}$, respectively.

We find that for the case $D=D_{\text{crit}}$,
there is a sign change of the Casimir force density, as shown in Fig. \ref{FofD2}.
That is because the zero temperature term of the Casimir force density is
\begin{equation}
	 F^{\text{Cas}.}_0=-\frac{D-p+1}{2^{D+2}L^{D-p+2}}\sum_{q=0}^{p-1}\frac{(-1)^qC_p^q\Gamma(\frac{D-q+1}2)Z_{p-q}(D-q+1)}{\pi^{\frac{D-q+1}2}},
	\label{F0}
\end{equation}
which has the same sign with $\varepsilon_0^{(\text{D}),\text{reg}.}$ expressed in eq.\eqref{e0DN2} for given $D,p$.
And the leading term of the high temperature expansion of the Casimir force density is
\begin{equation}
	F^{\text{Cas}.}_{T\rightarrow\infty}\sim-\frac{D-p}{2^{D+1}\beta L^{D-p}}\sum_{q=0}^{p-1}\frac{(-1)^qC_p^q\Gamma(\frac{D-q}2)Z_{p-q}(D-q)}{\pi^{\frac{D-q}2}},\quad T\rightarrow\infty.
	\label{Fasym}
\end{equation}
Comparing the two equations above one can find that for even $p$,
if $\varepsilon_0^{(\text{D}),\text{reg}.}$, and thus $F^{\text{Cas}.}_0$, change sign at $D=D_{\text{crit}}$\cite{caruso},
it will take $D=D_{\text{crit}}+1$ for $F^{\text{Cas}.}_{T\rightarrow\infty}$ to change sign.
That is, for even $p$ of Dirichlet case, at $D=D_{\text{crit}}$,
the zero temperature Casimir force density has changed to negative,
while the high temperature limit of the Casimir force density is still positive just as the case of $D<D_{\text{crit}}$.
So, there must be a cross of the Casimir force density from attractive to repulsive for $D=D_{\text{crit}}$.
This is also in accordance with Fig. \ref{p2d6DenergyL},
which shows the free energy density has a local maximum,
or i.e. a equilibrium point where the Casimir force density is zero.

In Fig. \ref{p2d7DforceL1}, we find that for Dirichlet BCs with even $p$ and $D>D_{\text{crit}}$,
the behavior of the Casimir force density as a function of the side length is generally similar to the odd $p$ cases or the other two BCs as shown in Fig. \ref{p2d3NforceL}.
Except that in Fig. \ref{p2d7DforceL2} we see a cross of the zero temperature line and the finite-temperature line.
This also can be seen in Fig. \ref{p2d7DforceT},
which shows that the absolute value of the Casimir force density decreases first and then increases as temperature rises.
Despite the little fluctuation with increase of temperature,
The Casimir force density is always negative in this case.

From Fig. \ref{FofN} - \ref{FofD3},
it is shown that the Casimir force density for Dirichlet BCs with even $p$ and $D<D_{\text{crit}}$ is repulsive,
for $D=D_{\text{crit}}$ it has a cross from attractive to repulsive with the inrease of temperature or the side length,
and it is attractive for other cases including Dirichlet BCs with even $p$ and $D>D_{\text{crit}}$ or odd $p$, Neumann BCs and periodic BCs.
The repulsive and the attractive forces (densities) are both increasing functions of temperature
and decreasing functions of the side length.
They vanish with the side length going to infinity.
\section{Conclusions and discussion}
\label{conclusion}
We have reconsidered the thermal scalar Casimir effect for $p$-dimensional rectangular cavity inside $D+1$-dimensional Minkowski space-time.
Our main  conclusions are as follows:

(i) We stressed that in order to get the reasonable physical result
one has to regularize both the zero temperature part and the temperature-dependent part of the free energy (density).
We provided a rigorous derivation of the regularization of the temperature-dependent part using Abel-plana formula
and obtained a general result that could include the cases of parallel plates and three-dimensional box in the literature.

(ii) For both $D>p$ and $D=p$ cases with periodic, Dirichlet and Neumann BCs,
we give the explicit expressions of the Casimir free energy (density) in both low temperature (small separations) and high temperature (large separations) regimes,
through which the asymptotic behavior of the free energy (density) changing with temperature and the side length is easy to see.

(iii) For $D>p$, for a given temperature, with the side length going to infinity,
the Casimir free energy density tends to a positive constant for Dirichlet BCs with odd $p$,
it tends to a negative constant for Dirichlet BCs with even $p$ and for Neumann BCs,
and it tends to zero for periodic BCs.
For Dirichlet BCs with even $p$ and $D<D_{\text{crit}}$,
the free energy (density) changes from positive to negative,
and it changes from negative to positive for Dirichlet BCs with odd $p$,
and for Neumann and periodic BCs both with $D=p$,
when changing with temperature or the side length.
But for other cases there is no such cross.

(iv) For $D=p$, the leading term of the Casimir free energy for all three BCs is a logarithmic function of the side length but not a constant as in $D>p$ case.

(v) The thermal Casimir force (density) for Dirichlet BCs with even $p$ and $D<D_{\text{crit}}$ is repulsive
and has a cross from attractive to repulsive with $D=D_\text{crit}$.
For the rest cases it is always attractive.
Both the repulsive and attractive forces (densities) are increasing functions of temperature and decreasing functions of the side length.
They vanish with the side length going to infinity.

   We hope we have provided a comprehensive and complete understanding to this old problem both analytically and numerically.
We argue that both the zero temperature and the temperature-dependent parts of the free energy (density) have to be regularized to get the physical result.
This is true for any configurations other than rectangular cavities when the temperature modification of the Casimir effect is considered.
Because the emphasize of this paper is the rigorous prove and derivation of the regularization
and obtain the physical results of the free energy and thermal Casimir force (density),
we have taken the $p$-dimensional cavity inside $D+1$-dimensional space-time as a hypercube,
ignoring the influence of the ratio of the side lengths,
which may affect the sign of the Casimir force as is known in zero temperature case.
In addition, the other thermodynamical quantities such as the entropy can also be considered
and the same problem for electromagnetic field is also worthy to study.

\section*{Acknowledgments}
This work is partially supported by the Key Project of
Chinese Ministry of Education.(No211059), Innovation Program of
Shanghai Municipal Education Commission(11zz123) and Program of Shanghai Normal University (DXL124).


\end{document}